\documentclass[twocolumn,showpacs,prl]{revtex4}
\usepackage{graphicx}

\begin{document}

\title{Single-Walled Carbon Nanotubes as Shadow Masks for Nanogap Fabrication}
\author{E.\ P.\ De Poortere$^a$ and H.\ L.\ Stormer$^a$}
\address{Department of Physics, Columbia University, New York, NY 10027}
\author{L.\ M.\ Huang$^a$, S.\ J.\ Wind$^a$, and S.\ O'Brien$^a$}
\address{Department of Applied Physics and Applied Mathematics, Columbia University, New York, NY 10027}
\author{M. Huang$^a$, and J.\ Hone$^a$}
\address{Department of Mechanical Engineering, Columbia University, New York, NY 10027 }

\date{\today}

\begin{abstract}
We describe a technique for fabricating nanometer-scale gaps in Pt
wires on insulating substrates, using individual single-walled
carbon nanotubes as shadow masks during metal deposition. More than
80 \% of the devices display current-voltage dependencies
characteristic of direct electron tunneling. Fits to the
current-voltage data yield gap widths in the 0.8 -- 2.3 nm range for
these devices, dimensions that are well suited for single-molecule
transport measurements.
\end{abstract}

 \maketitle

Electronic devices based on single aromatic molecules have raised
considerable interest in the last few years, as a variety of new
techniques now make it possible to fabricate metal leads --
separated by only a few nanometers -- that can be bridged by a
single molecule \cite{ebeam, park99, reichert02}. Nevertheless, in
spite of considerable improvements, these popular nanofabrication
techniques produce gap junctions that are relatively rough at the nm
scale. Electromigration, for example, is known to give rise to
fairly disordered gaps containing metal islands or clusters
\cite{hatzor05}, which make conductance data difficult to interpret.
This issue imposes serious limits on interpreting current-voltage
data obtained with such devices, as single-molecule transport
experiments ideally require a large number of clean and similar
metal junctions, and electron transport through organic molecules
appears to be extremely sensitive to the geometry of the
molecule-metal contact \cite{ratner05}.

In this Letter we report a nanogap fabrication technique based on a
single-walled nanotube (SWNT) shadow mask, in which the dimension of
the junction is controlled solely by the diameter of the nanotube
and by its distance from the substrate during metal deposition. This
scheme has been applied in the past with thicker multi-walled carbon
nanotubes or SWNT bundles \cite{lefebvre00} to create 20 nm-wide
gaps. Here we rely on a mechanical transfer technique developed by
Huang {\it et al.} \cite{xmhhuang05} to place long, quasi parallel,
individual SWNT's onto a surface patterned with an electron-beam
resist film. In this way, a given nanotube is suspended over up to
ten parallel ``trenches'' in the patterned resist; after metal
deposition and lift-off, the method thus gives rise to ten nanoscale
gaps {\it defined by the same nanotube}. We measure the
current-voltage characteristics of 79 wires produced with this
method, and deduce from tunneling data that 66 of these have widths
in the 0.8 -- 2.3 nm range. Only two out of 79 wires are shorted.

\begin{figure}
\includegraphics[scale=0.6]{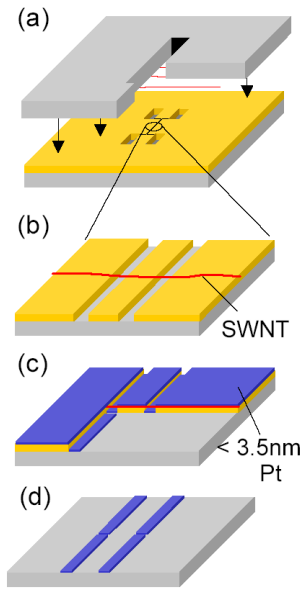}
\caption{Fig. 1. Sample fabrication steps. (a) Single-walled
nanotubes are transferred by pressing the carrier chip onto the
resist-patterned sample. SWNT sections that were suspended on the
carrier chip adhere preferentially to the resist and break off from
the rest of the original nanotube. For clarity, only two trenches
are shown, and at a larger scale; actual device contains twelve sets
of ten wires each. (b) Detail of central area of the sample: SWNT
now rests over lithographically-defined trenches in the 35 nm-thick
copolymer/PMMA resist. (c) Up to 3.5 nm Pt are deposited over the
sample, forming a wire and a nm-scale gap opening underneath the
overhanging nanotube. (d) Excess metal is removed after lift-off.}
\label{scheme}
\end{figure}
The metal and substrate materials we choose for wire fabrication are
Pt and ZrO$_2$, respectively. This choice is inspired by reports
that the growth of Pt films over clean cubic ZrO$_2$(100) crystals
is two-dimensional \cite{roberts91,dilara95}, in contrast to the
typical grain-like growth of noble metals on oxides
\cite{campbell97}. This layer-by-layer film growth allows us, in
principle, to work with metal films much thinner than those used
traditionally for this type of device, which are typically about
tens of nm thick \cite{PtonSi}. Using ultrathin metal films allows
us in turn to reduce the thickness of the resist film, and therefore
to reduce the nm-scale broadening of features caused by the finite
size of the metal source during Pt evaporation. Very importantly,
thin Pt films also result in shallower junctions whose structure and
contents could be observed by scanning probes.

The devices are made from n-doped Si substrates, coated with 3.7 nm
or 9.1 nm-thick ZrO$_2$ films formed by atomic layer deposition
(ALD). We first use standard electron-beam lithography to pattern
long, narrow trenches in a resist bilayer \cite{resist}, developed
in a 1:3 H$_2$O:isopropanol cosolvent at $\sim$ 4$^o$C with
ultrasonic agitation. On a separate Si ``carrier'' chip, long SWNT's
are grown by chemical vapor deposition, using ethanol as carbon
feedstock, according to a technique detailed in \cite{xmhhuang05}.
These wafers contain 100 $\mu$m x 1 mm window slits over which
typically 4--10 carbon nanotubes rest suspended after growth.

We then transfer the nanotubes from their carrier chip onto the
Si/ZrO$_2$/resist sample by bringing the two samples into contact
\cite{xmhhuang05} [Fig.\ \ref{scheme}(a)]. The suspended part of the
nanotubes on the carrier chip adheres to the resist on the
Si/ZrO$_2$ sample, and remains on it when the two samples are
separated, breaking away from the other nanotube sections [Fig.\
\ref{scheme}(b)]. The slit on the carrier sample can be easily
aligned with the resist trenches under the microscope, so that the
nanotubes, once transferred on the resist, are now suspended above
the trenches. This method allows us currently to transfer 1--4
nanotubes from a carrier chip, though this number can probably be
increased by minimizing particulate density on the samples or by
reducing the total contact area between carrier and resist samples
\cite{bundles}.

Following the nanotube transfer, we deposit $\sim$ 3.5 nm Pt onto
the patterned sample by electron-beam evaporation at 0.2 -- 0.4
\AA/s. We note that the design of the evaporation chamber and of the
metal source are critical for nm-scale gap fabrication, since wide
sources or short evaporation distances can taper the edges of the
deposited metal by a few nm. Our Pt source is about 5 mm wide, and
the source-to-sample distance 75 cm in our process. Given these
dimensions, the nanotube shadow width may be reduced by up to 0.2 nm
\cite{lefebvre00}, much less than the diameter of the smallest SWNT.
After metal deposition, the sample is placed in boiling acetone for
two hours and dipped in an ultrasound bath for 30 s in the same
solvent. These steps ensure the removal of both resist and
nanotubes, as well as that of the metal in the unpatterned areas.

A potential problem with this technique is that of nanotube sagging,
which can cause the tubes to adhere to the oxide surface exposed at
the bottom of the trenches, and ultimately prevent these tubes from
lifting off properly after metal evaporation. We designed the
trenches to be 35 nm deep and 40 -- 200 nm wide, and were able to
lift off easily all nanotubes transferred on these samples,
indicated that this is not an issue for our samples.

\begin{figure}
\includegraphics[scale=.5]{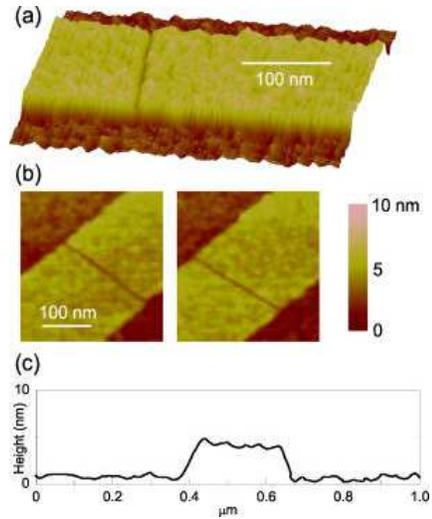}
\caption{Fig. 2. (a) and (b) AFM scans of device, showing a gap in
the Pt metal line formed by the SWNT shadow mask. (b) Scans of two
different wires with gaps, produced using the \emph{same nanotube}.
(c) Transversal linescan of a Pt wire, showing that the roughness of
the Pt surface and of the ZrO$_2$ substrate are similar (the RMS
roughness for both is $\sim$ 0.35 nm).} \label{scans}
\end{figure}
Figure \ref{scans} shows atomic force microscope (AFM) scans of some
of our devices. A narrow, continuous groove cuts across the Pt wire,
a result of the shadow evaporation of the metal around the suspended
nanotube. The gap width is difficult to determine precisely from AFM
scans because of the finite size of the scanning tip, but can be
inferred from measurements of the tunneling current across the gap,
as detailed below, and shown to be smaller than $\sim$ 2.5 nm.
Several gaps are produced in this fashion from each nanotube,
suspended across a sequence of parallel trenches. The resulting gaps
are shown for two Pt wires in Fig.\ \ref{scans}(b).

The linescan of the metal line [Fig. \ref{scans}(c)] shows that the
apparent roughness of the Pt film is comparable to that of the
underlying oxide, indicating that the film, in spite of its small
thickness, can form a continuous layer on ZrO$_2$. Film continuity
is also confirmed by conductance measurements, discussed below.

\begin{figure}
\includegraphics[scale=1.1]{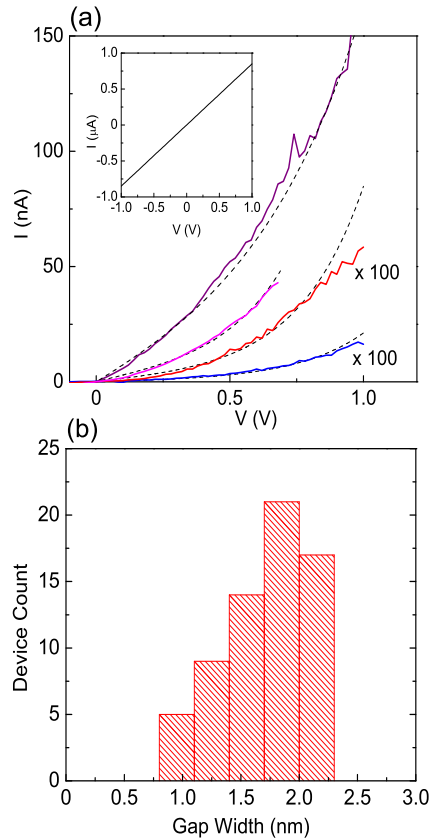}
\caption{Fig. 3. (a) Current-voltage data for four nanogaps, showing
nonlinear dependence characteristic of direct tunneling. The devices
measured here were obtained from four parallel wires ``cut'' by the
same nanotube. Dashed curves: calculated fits to the data using the
Simmons model \cite{simmons63}. Calculated gap widths are, from top
to bottom, 1.3, 1.8, and 2 nm for the last two curves. Inset: I-V
data for a 3.5 nm-thick Pt wire (40 nm $\times$ 90 $\mu$m) with no
nanogap. (b) Histogram of gap widths obtained from I-V tunneling
data for 66 different wires. Most of the widths fall within the 1--2
nm range.} \label{IV}
\end{figure}

We study electrical transport through these wires and gaps in air
and at room temperature. Wire thicknesses are about 3.5 nm for three
of our devices, and about 2.5 nm for the fourth one, and the uncut
wires we measured are $\sim$ 90 $\mu$m long and 40 -- 100 nm wide.
3.5 nm-thick wires have a total resistance of 0.6 -- 1.2 M$\Omega$,
i.e., $\sim$ 700 $\Omega$/square [see inset of Fig.\ \ref{IV}(a)]
\cite{2nm}. We note that these resistances, although in the
M$\Omega$ range, are small relative to typical resistances of
metal-molecule-metal bridges \cite{salomon03}, and can be further
reduced by appropriate design. We then study a total of 79 wires
interrupted by nanogaps, and show data for four of these in Fig.\
\ref{IV}(a). 66 of these wires show electron transport
characteristic of direct tunneling, while the I-V traces are linear
-- and hence the gaps shorted -- in only two out of the 79 wires.
The remaining 11 gaps carry a current smaller than about 10 pA at 2
V bias. In contrast, most wires produced during the same process,
but not shadowed by SWNT's, do not show tunneling I-V's.

We analyze the I-V's of all 66 ``tunneling'' wires by fitting our
data to Simmons' model \cite{simmons63}, taking into account the
influence of the image charge within the gap. In this model, gap
width, barrier height and tunneling area are fitting parameters
\cite{modelnote}. Data and calculated curves for some of the devices
are plotted in Fig.\ \ref{IV}(a). We obtain gap widths ranging from
0.8 to 2.3$\pm$0.2 nm, with a median gap width smaller than 2 nm
\cite{barrierheight}. A histogram of these data is shown in Fig.\
\ref{IV}(b). We note that these widths are consistent with the
measured diameter range of the nanotubes used in our process, 0.8 nm
-- 1.8 nm, obtained from Raman scattering data \cite{lhuang05}.

In summary, we have developed a technique for creating thin
nanoscale metal junctions, whose widths are controlled by the
diameter of a SWNT shadow mask. The yield of gaps in the tunneling
range (0.8 -- 2.3 nm here) is larger than 80\%, and less than 3\% of
the devices are shorted, indicating that the gaps are likely free of
metal clusters. In addition, this technique helps to avoid e-beam
resist contamination issues at the metal gap, a potential problem
with direct electron-beam lithography. These devices are thus well
suited for single-molecule transport measurements, which require
excellent control over gap fabrication at the nanometer scale.
Finally, the 2.5 --  3.5 nm metal thickness used in our process
opens up the possibility of imaging with scanning probes
single-molecule devices based on these metal films.

We thank Evgeni Gusev for the ALD growth, and Philip Kim for access
to the AFM. We also thank Henry Huang for helpful discussions about
the nanotube transfer technique. This work was supported by the
Nanoscale Science and Engineering Initiative of the National Science
Foundation under NSF Award Number CHE-0117752, by the NSF grant
PHY-0103552, by the New York State Office of Science, Technology,
and Academic Research (NYSTAR), and by a grant from the W.\ M.\ Keck
Foundation.

$^a$Center for Electron Transport in Molecular Nanostructures,
Columbia University, New York, NY 10027


\begin{references}
\bibitem{resist}
The lower layer is a MMA/8.5/MAA EL1.5 copolymer, and the upper
layer 495K PMMA A1. Total resist thickness is about 35 nm.
\bibitem{2nm}
The 2.5 nm-thick wires were approximately six times more resistive
than the 3.5 nm-thick ones, although the former were measured after
an acryl strip cleaning step, which likely oxidizes part of the Pt.
\bibitem{modelnote}
In principle, electrons tunnel through both air and the underlying
ZrO$_2$ substrate, though additional simulations show that the oxide
carries a much smaller direct tunneling current than air. In
practice, we simplify the model by assuming that only one
``effective'' material, with a relative permittivity equal to 1,
separates the electrodes.
\bibitem{barrierheight}
The barrier height parameters that best fit our data vary between 4
and 6 eV. This energy variation probably results from the
sensitivity of the metal work function to adsorbed gases.
\end{references}
\end{document}